\documentclass[10pt]{article}

\usepackage{graphicx}

\usepackage[all,cmtip]{xy}

\usepackage{setspace}
\usepackage{bbm,youngtab}
\usepackage{amscd,mathrsfs}
\usepackage{amssymb,amsmath,dsfont,amsfonts,amsthm}
\usepackage{euscript,enumerate}
\usepackage{stmaryrd}
\usepackage[cbgreek]{textgreek} 
\usepackage{booktabs}
\usepackage{multirow}
\usepackage{hhline} 
\usepackage{color}

\DeclareMathAlphabet{\mathpzc}{OT1}{pzc}{m}{it}

\allowdisplaybreaks 

\usepackage{authblk}
\usepackage{natbib}
\usepackage{color, soul}
\usepackage{enumerate}
\usepackage{empheq}
\usepackage{rotating}
\usepackage{ftnxtra}
\usepackage{fnpos}
\usepackage[titletoc,toc,title]{appendix}
\usepackage{euscript}
\usepackage{graphicx}
\usepackage{epsfig}
\usepackage{epstopdf}
\DeclareGraphicsExtensions{.pdf,.png,.jpg,.eps}
\usepackage{pstool}
\usepackage{upgreek}
\usepackage{mathrsfs}
\usepackage{tikz-cd}

\usepackage{psfrag}

\numberwithin{equation}{section}

\theoremstyle{plain}	
\newtheorem{thm}{Theorem}[section]

\newtheorem{prop}[thm]{Proposition}
\newtheorem*{prop*}{Proposition}
\theoremstyle{definition}	

\newtheorem{remark}[thm]{Remark}

\usepackage{caption}
\usepackage{subcaption}

\usepackage{hyperref}
\hypersetup{colorlinks=true, linkcolor=blue}
\hypersetup{colorlinks=true,citecolor=blue}

\DeclareMathAlphabet{\mathpzc}{OT1}{pzc}{m}{it}

\usepackage{amsmath, amsthm, amssymb}

\usepackage{cleveref}

\DeclarePairedDelimiter\abs{\lvert}{\rvert}

\makeatletter
\newsavebox{\@brx}
\newcommand{\llangle}[1][]{\savebox{\@brx}{\(\m@th{#1\langle}\)}%
  \mathopen{\copy\@brx\mkern2mu\kern-0.9\wd\@brx\usebox{\@brx}}}
\newcommand{\rrangle}[1][]{\savebox{\@brx}{\(\m@th{#1\rangle}\)}%
  \mathclose{\copy\@brx\mkern2mu\kern-0.9\wd\@brx\usebox{\@brx}}}%
\let\oldabs\abs
\def\abs{\@ifstar{\oldabs}{\oldabs*}}
\makeatother

\usepackage{accents}

\usepackage{mathtools}

    {\end{bmatrix}}%

\usepackage[all,cmtip]{xy}
\usepackage{bm}

\newcommand{\sigmac}{\accentset{c}{\sigma}}
\newcommand{\sigmar}{\accentset{r}{\sigma}}

\usepackage{enumitem}

\begin{document}

\title{\textbf{Universal Displacements in Inextensible Fiber-Reinforced Linear Elastic Solids}}

\author[1,2]{Arash Yavari\thanks{Corresponding author, e-mail: arash.yavari@ce.gatech.edu}}
\affil[1]{\small \textit{School of Civil and Environmental Engineering, Georgia Institute of Technology, Atlanta, GA 30332, USA}}
\affil[2]{\small \textit{The George W. Woodruff School of Mechanical Engineering, Georgia Institute of Technology, Atlanta, GA 30332, USA}}

\maketitle

\begin{abstract}
For a given class of materials, universal displacements are those displacements that can be maintained for any member of the class by applying only boundary tractions. In this paper we study universal displacements in compressible anisotropic linear elastic solids reinforced by a family of inextensible fibers. For each symmetry class and for a uniform distribution of straight fibers respecting the corresponding symmetry we characterize the respective universal displacements. A goal of this paper is to investigate how an internal constraint affects the set of universal displacements. We have observed that other than the triclinic and cubic solids in the other five classes (a fiber-reinforced solid with straight fibers cannot be isotropic) the presence of inextensible fibers enlarges the set of universal displacements.
\end{abstract}

\begin{description}
\item[Keywords:] Universal displacement, universal deformation, fiber-reinforced solids, anisotropic solids.
\end{description}

\tableofcontents

\section{Introduction}

A universal motion (deformation or displacement) is one that can be maintained in the absence of body forces for all materials in some given class. In other words, a universal motion of a body can be maintained by applying only boundary tractions when the body is made of any material in the given class, e.g., homogeneous compressible isotropic solids, or homogeneous incompressible isotropic solids. In nonlinear elasticity universal motions have been important both experimentally \citep{Rivlin1951} and theoretically \citep{Tadmor2012,Goriely2017}. The notion of universal deformations was introduced in the two seminal papers of Jerry Ericksen \citep{Ericksen1955,Ericksen1954}. \citet{Ericksen1955} showed that for homogenous compressible isotropic solids universal deformations are homogeneous. Ericksen's study of universal deformations in homogeneous incompressible isotropic solids \citep{Ericksen1954} was motivated by some earlier works of Ronald Rivlin \citep{Rivlin1948, Rivlin1949a, Rivlin1949b}. The characterization of universal deformations in the presence of internal constraints turns out to be a more difficult problem \citep{Saccomandi2001}. \citet{Ericksen1954} found four families of universal deformations for incompressible isotropic elastic solids. Later on a fifth family of universal deformations was discovered \citep{SinghPipkin1965,KlingbeilShield1966}. \citet{Ericksen1954} had conjectured that a deformation with constant principal invariants is homogeneous, and this turned out to be incorrect \citep{Fosdick1966}. The universal deformations in the fifth family have constant principal invariants but are not homogeneous. To this date, it is not known if there are other inhomogeneous constant-principal invariant universal deformations.

There have been recent extensions of Ericksen's analysis to inhomogeneous isotropic (both compressible and incompressible) elasticity \citep{Yavari2021}, anisotropic elasticity \citep{YavariGoriely2021,Yavari2022Universal}, and anelasticity \citep{YavariGoriely2016,Goodbrake2020}. The analogue of universal deformations in linear elasticity are universal displacements \citep{Truesdell1966,Gurtin1972,Yavari2020}. For compressible anisotropic linear elastic solids the universal displacements were characterized for all the eight anisotropy classes in \citep{Yavari2020}. In particular, it was shown that the larger the symmetry group the larger the set of universal displacements. Thus, isotropic solids have the largest set of universal displacements while triclinic solids have the smallest set of universal displacements. The analysis of universal displacements was recently extended to inhomogeneous solids \citep{YavariGoriely2022}, and to linear anelasticity \citep{Yavari2022Anelastic-Universality}.

A class of solids with internal constraints that have important engineering applications are materials  reinforced with inextensible fibers \citep{Pipkin1971,Pipkin1974,Pipkin1979}.
There are very few works on universal deformations of fiber-reinforced solids in the literature. 
\citet{Beskos1972} considered homogeneous compressible isotropic solids reinforced with inextensible fibers and investigated the possibility of the universal deformations of incompressible isotropic solids being universal for this class of solids as well. More specifically, Families $1$, $2$, $3$, and $4$ were examined. It was shown that some subsets of these families are universal for certain fiber distributions. All these universal deformations turn out to be homogeneous with the exception of shearing of a circular tube with circumferential fibers.
\citet{Beatty1978} considered homogenous compressible isotropic solids reinforced by one family of inextensible fibers. He investigated the problem of finding all those fiber distributions for which homogenous deformations are universal. He showed that there are only three types of such fiber distributions. In all the three distributions fibers are straight lines.

In this paper, we study universal displacements in compressible anisotropic linear elastic solids reinforced with one family of inextensible straight fibers. For each symmetry class we characterize the set of universal displacements and compare it with that of compressible solids without reinforcement.

This paper is organized as follows. In \S\ref{Sec:LinearElasticity} we briefly review linear elasticity in the presence of internal constraints. The constitutive and equilibrium equations of anisotropic compressible linear elastic solids reinforced by a family of inextensible fibers are discussed in \S\ref{Sec:Fiber-Reinforced}. In \S\ref{Sec:Universal-Displacements}, the universal displacements of each class of fiber-reinforced solids (triclinic, monoclinic, tetragonal, trigonal, orthotropic, transversely isotropic, and cubic) are characterized.
Conclusions are given in \S\ref{Sec:Conclusions}.

\section{Linear elasticity of materials with internal constraints} \label{Sec:LinearElasticity}

In this section we review the governing equations of linear elasticity with internal constraints. 
Let us consider a body $\mathcal{B}$ that is made of a linear elastic solid and has an internal constraint at every point $\mathbf{x}\in\mathcal{B}$.
It is assumed that at $\mathbf{x}\in\mathcal{B}$, the displacement gradient satisfies the following linear constraints \citep{Pipkin1976}
\begin{equation} \label{Constraints}
    \mathfrak{c}^b{}_a\,u^a{}_{|b}=0\,,
\end{equation}
where $\mathfrak{c}^b{}_a=\mathfrak{c}^b{}_a(\mathbf{x})$, and it is assume that $\mathfrak{c}^{ab}=\mathfrak{c}^{ba}$. 
Thus, \eqref{Constraints} can be rewritten in terms of strain as $\mathfrak{c}^{ab}\,\epsilon_{ab}=0$.
In a coordinate-free form the internal constraint is written as $\operatorname{tr}(\boldsymbol{\mathfrak{c}}\cdot\nabla\mathbf{u})=0$, or $\operatorname{tr}(\boldsymbol{\mathfrak{c}}\cdot\boldsymbol{\epsilon})=0$.  In the presence of an internal constraint the stress has a constitutive part and a reactive part: $\sigma^{ab}=\sigmac^{ab}+\sigmar^{ab}$ such that $\sigmar^{ab}\epsilon_{ab}=0$.
The stress-strain relationship is written as $\epsilon_{ab}=c_{abcd}\,\sigma^{cd}$, where $c_{abcd}$ are components of the compliance tensor. The internal constraint can be rewritten as $\mathfrak{c}^{ab}\,c_{abcd}\,\sigma^{cd}=0$, and this must hold for any stress $\sigma^{cd}$. Thus, $\mathfrak{c}^{ab}\,c_{abcd}=0$, which are six constraints on the compliance components. Hence, internal constraints reduce the number of independent compliance components (from $21$ to $15$), and consequently, the number of independent elastic constants. 
In Voigt notation the bijection $(11,22,33,23,13,12)\leftrightarrow(1,2,3,4,5,6)$ is used.
The stress-strain relationship is rewritten as $\epsilon_{\alpha}=c_{\alpha\beta}\,\sigma_{\beta}$ (with Greek indices running from 1 to 6), where the compliance tensor is represented by a symmetric $6\times 6$ stiffness matrix $\mathbf{c}$:
\begin{equation} 
    \mathbf{c}=\begin{bmatrix}
    c_{11} & c_{12} & c_{13} & c_{14} & c_{15} & c_{16} \\
    c_{12} & c_{22} & c_{23} & c_{24} & c_{25} & c_{26}  \\
    c_{13} & c_{23} & c_{33} & c_{34} & c_{35} & c_{36}  \\
    c_{14} & c_{24} & c_{34} & c_{44} & c_{45} & c_{46}  \\
    c_{15} & c_{25} & c_{35} & c_{45} & c_{55} & c_{56}  \\
    c_{16} & c_{26} & c_{36} & c_{46} & c_{56} & c_{66}  
 \end{bmatrix}\,.
\end{equation}
The internal constraint $\mathfrak{c}^{ab}\,\epsilon_{ab}=0$ in Voigt notation is written as $\mathfrak{c}_{\alpha}\,\epsilon_{\alpha}=\mathfrak{c}_{\alpha}\,c_{\alpha\beta}\,\sigma_{\beta}=0$, and hence, $c_{\alpha\beta}\,\mathfrak{c}_{\alpha}=0$, which implies that $\boldsymbol{\mathfrak{c}}$ is an eigenvector of $\mathbf{c}$ with a zero eigenvalue \citep{Pipkin1976}. If there is only one internal constraint, the other eigenvalues of $\mathbf{c}$ are assumed to be nonzero.
The reaction stress $\sigmar_{\alpha}$ satisfies the relation $\sigmar_{\alpha}\,\epsilon_{\alpha}=0$, and thus, $\sigmar_{\alpha}=\Lambda \mathfrak{c}_{\alpha}$, where $\Lambda$ is constitutively indeterminate and is determined after solving the equilibrium equations and enforcing the boundary conditions.
Let us consider a compressible solid reinforced by one family of inextensible fibers. At $\mathbf{x}\in\mathcal{B}$, let us denote the unit tangent to the fiber by $\mathbf{n}(\mathbf{x})$. Thus, $n^a n^b\,g_{ab}=1$. Inextensibility is equivalent to $\epsilon_{ab}\,n^an^b=0$, or $u_{a,b}\,n^an^b=0$. Thus, in this case, $\mathfrak{c}^{ab}=n^a n^b$. The stress tensor is written as $\sigma^{ab}=T n^a n^b+s^{abcd}\,\epsilon_{cd}$, where $T=T(\mathbf{x})$ is an unknown tension field, and $s^{abcd}$ are the components of the elasticity (stiffness) tensor \citep{Adkins1955,Truesdell2004,Saccomandi2002}.

\section{Fiber-reinforced anisotropic linear elastic solids}  \label{Sec:Fiber-Reinforced}

Let us consider a compressible anisotropic linear elastic solid that is reinforced by a family of inextensible fibers. We assume a uniform distribution of fibers parallel to the $x_3$-axis in a Cartesian coordinate system $(x_1,x_2,x_3)$. 

\paragraph{Elastic constants.}
The constraints $\mathfrak{c}^{ab}\,c_{abcd}=0$ imply that $c_{33cd}=0$, which in Voigt notation are written as $c_{3\beta}=c_{\beta 3}=0$, $\beta=1,\hdots,6$, i.e., the compliance matrix has the following form:
\begin{equation} \label{Compliance-1Fiber}
    \mathbf{c}=\begin{bmatrix}
    c_{11} & c_{12} & 0 & c_{14} & c_{15} & c_{16} \\
    c_{12} & c_{22} & 0 & c_{24} & c_{25} & c_{26}  \\ 
    0 & 0 & 0 & 0 & 0 & 0  \\
    c_{14} & c_{24} & 0 & c_{44} & c_{45} & c_{46}  \\
    c_{15} & c_{25} & 0 & c_{45} & c_{55} & c_{56}  \\
    c_{16} & c_{26} & 0 & c_{46} & c_{56} & c_{66}  
 \end{bmatrix}\,.
\end{equation}
Obviously, $\mathbf{c}$ is singular and cannot be directly inverted to find the matrix of elastic constants $\mathbf{s}=[s_{\alpha\beta}]$. 
The eigenvector $\mathfrak{c}_{\alpha}$ has components $\left[ 0 ~ 0 ~ 1 ~ 0 ~ 0 ~ 0\, \right]^{\mathsf{T}}$, and hence, stress can be written as
\begin{equation} 
	\sigma_{\alpha}=\sum_{\beta=1}^{6}s_{\alpha\beta}\,\epsilon_{\beta}+T\delta_{\alpha 3}
	=\sum_{\beta=1}^{6}\sum_{\gamma=1}^{6}s_{\alpha\beta}\,c_{\beta\gamma}\,\sigma_{\gamma}
	+T\delta_{\alpha 3}
	=\sum_{\substack{\gamma=1 \\ \gamma\neq 3}}^6 \sum_{\substack{\beta=1 \\ \beta\neq 3}}^6
	s_{\alpha\beta}\,c_{\beta\gamma}\,\sigma_{\gamma}
	+T\delta_{\alpha 3}\,.
\end{equation}
For $\alpha\neq 3$, this is written as
\begin{equation} 
	\sigma_{\alpha}=\sum_{\substack{\gamma=1 \\ \gamma\neq 3}}^6
	\left(\sum_{\substack{\beta=1 \\ \beta\neq 3}}^6 s_{\alpha\beta}\,c_{\beta\gamma}
	\right)\sigma_{\gamma} \,,
\end{equation}
and hence
\begin{equation} \label{c-s-relation}
	\sum_{\substack{\beta=1 \\ \beta\neq 3}}^6
	s_{\alpha\beta}\,c_{\beta\gamma}=\delta_{\alpha\gamma} 
	\,,\quad \alpha\,,\gamma =1,2,4,5,6\,.
\end{equation}
This implies that the $5\times 5$ matrix obtained from $\mathbf{c}$ by removing its third row and column can be inverted to give the corresponding $5\times 5$ elasticity (stiffness) matrix, i.e.,
\begin{equation} \label{reduced-c-triclinic}
	\bar{\mathbf{s}}=\begin{bmatrix}
    s_{11} & s_{12} &  s_{14} & s_{15} & s_{16} \\
    s_{12} & s_{22} & s_{24} & s_{25} & s_{26}  \\ 
    s_{14} & s_{24}  & s_{44} & s_{45} & s_{46}  \\
    s_{15} & s_{25}  & s_{45} & s_{55} & s_{56}  \\
    s_{16} & s_{26}  & s_{46} & s_{56} & s_{66}
     \end{bmatrix}=
 \begin{bmatrix}
    c_{11} & c_{12} &  c_{14} & c_{15} & c_{16} \\
    c_{12} & c_{22} & c_{24} & c_{25} & c_{26}  \\ 
    c_{14} & c_{24}  & c_{44} & c_{45} & c_{46}  \\
    c_{15} & c_{25}  & c_{45} & c_{55} & c_{56}  \\
    c_{16} & c_{26}  & c_{46} & c_{56} & c_{66}  
 \end{bmatrix}^{-1}=\bar{\mathbf{c}}^{-1}\,.
\end{equation}
We call $\bar{\mathbf{c}}$ and $\bar{\mathbf{s}}$ the reduced compliance and stiffness matrices, respectively.
For $\alpha= 3$, 
\begin{equation} 
	\sigma_3=\sum_{\substack{\gamma=1 \\ \gamma\neq 3}}^6 
	\sum_{\substack{\beta=1 \\ \beta\neq 3}}^6 s_{3\beta}\,c_{\beta\gamma}\,\sigma_{\gamma}+T\,.
\end{equation}
As $T$ is an unknown, one can simply write $\sigma_3=T$ \citep{Hayes1975}. Thus, the stress-strain relationship is written as
\begin{equation} 
    \begin{bmatrix}
    \sigma_1 \\ 
    \sigma_2 \\ 
    \sigma_3 \\ 
    \sigma_4 \\ 
    \sigma_5 \\ 
    \sigma_6 
 \end{bmatrix}
    =\begin{bmatrix}
    s_{11} & s_{12} & 0 & s_{14} & s_{15} & s_{16} \\
    s_{12} & s_{22} & 0 & s_{24} & s_{25} & s_{26}  \\ 
    0 & 0 & T & 0 & 0 & 0  \\
    s_{14} & s_{24} & 0 & s_{44} & s_{45} & s_{46}  \\
    s_{15} & s_{25} & 0 & s_{45} & s_{55} & s_{56}  \\
    s_{16} & s_{26} & 0 & s_{46} & s_{56} & s_{66}  
 \end{bmatrix}
 \begin{bmatrix}
    \epsilon_1 \\ 
    \epsilon_2 \\ 
    1 \\ 
    \epsilon_4 \\ 
    \epsilon_5 \\ 
    \epsilon_6 
 \end{bmatrix}
 \,.
\end{equation}


\paragraph{Equilibrium equations.}
In the case of a homogeneous compressible anisotropic linear elastic solid, and in the absence of body forces the equilibrium equations with respect to a Cartesian coordinate system $(x_1,x_2,x_3)$ read $\sigma_{ab,b}=(c_{abcd}\epsilon_{cd})_{,b}=c_{abcd}\epsilon_{cd,b}=0$. In the Voigt notation this is rewritten as
\begin{equation} 
    \begin{bmatrix}
    \frac{\partial}{\partial x_1} & 0 & 0 & 0 & \frac{\partial}{\partial x_3} & \frac{\partial}{\partial x_2} \\
    0 & \frac{\partial}{\partial x_2} & 0 & \frac{\partial}{\partial x_3} & 0 & \frac{\partial}{\partial x_1} \\
    0 & 0 & \frac{\partial}{\partial x_3} & \frac{\partial}{\partial x_2}  & \frac{\partial}{\partial x_1} & 0
    \end{bmatrix}\begin{bmatrix}
    s_{11} & s_{12} & s_{13} & s_{14} & s_{15} & s_{16} \\
    s_{12} & s_{22} & s_{23} & s_{24} & s_{25} & s_{26}  \\
    s_{13} & s_{23} & s_{33} & s_{34} & s_{35} & s_{36}  \\
    s_{14} & s_{24} & s_{34} & s_{44} & s_{45} & s_{46}  \\
    s_{15} & s_{25} & s_{35} & s_{45} & s_{55} & s_{56}  \\
    s_{16} & s_{26} & s_{36} & s_{46} & s_{56} & s_{66}  
    \end{bmatrix}
    \begin{bmatrix}
    \frac{\partial u_1}{\partial x_1}  \\
    \frac{\partial u_2}{\partial x_2}  \\
    \frac{\partial u_3}{\partial x_3}  \\
    \frac{\partial u_2}{\partial x_3}+\frac{\partial u_3}{\partial x_2}  \\
    \frac{\partial u_1}{\partial x_3}+\frac{\partial u_3}{\partial x_1}  \\
    \frac{\partial u_1}{\partial x_2}+\frac{\partial u_2}{\partial x_1}   
    \end{bmatrix}=
    \begin{bmatrix}
    0 \\ 0 \\ 0   
    \end{bmatrix}\,.
\end{equation}
For the fiber-reinforced solid the above equilibrium equations are modified to read
\begin{equation} \label{Equilibrium-Equations}
    \begin{bmatrix}
    \frac{\partial}{\partial x_1} & 0 & 0 & 0 & \frac{\partial}{\partial x_3} & \frac{\partial}{\partial x_2} \\
    0 & \frac{\partial}{\partial x_2} & 0 & \frac{\partial}{\partial x_3} & 0 & \frac{\partial}{\partial x_1} \\
    0 & 0 & \frac{\partial}{\partial x_3} & \frac{\partial}{\partial x_2}  & \frac{\partial}{\partial x_1} & 0
    \end{bmatrix}
    \begin{bmatrix}
    s_{11} & s_{12} & 0 & s_{14} & s_{15} & s_{16} \\
    s_{12} & s_{22} & 0 & s_{24} & s_{25} & s_{26}  \\ 
    0 & 0 & T & 0 & 0 & 0  \\
    s_{14} & s_{24} & 0 & s_{44} & s_{45} & s_{46}  \\
    s_{15} & s_{25} & 0 & s_{45} & s_{55} & s_{56}  \\
    s_{16} & s_{26} & 0 & s_{46} & s_{56} & s_{66}  
 \end{bmatrix}
    \begin{bmatrix}
    \frac{\partial u_1}{\partial x_1}  \\
    \frac{\partial u_2}{\partial x_2}  \\
    1  \\
    \frac{\partial u_2}{\partial x_3}+\frac{\partial u_3}{\partial x_2}  \\
    \frac{\partial u_1}{\partial x_3}+\frac{\partial u_3}{\partial x_1}  \\
    \frac{\partial u_1}{\partial x_2}+\frac{\partial u_2}{\partial x_1}   
    \end{bmatrix}=    
    \begin{bmatrix}
    0 \\ 0 \\ 0   
    \end{bmatrix}\,.
\end{equation}
Notice that the third equilibrium equation (along with the traction boundary conditions) determines the tension field $T=T(x_1,x_2,x_3)$. The remaining two equilibrium equations are written as
\begin{equation} \label{Universality-Inextensible}
    \begin{bmatrix}
    \frac{\partial}{\partial x_1} & 0 & 0 & \frac{\partial}{\partial x_3} & \frac{\partial}{\partial x_2} \\
    0 & \frac{\partial}{\partial x_2} & \frac{\partial}{\partial x_3} & 0 & \frac{\partial}{\partial x_1} 
    \end{bmatrix}
    \begin{bmatrix}
    s_{11} & s_{12} & s_{14} & s_{15} & s_{16} \\
    s_{12} & s_{22} & s_{24} & s_{25} & s_{26}  \\ 
    s_{14} & s_{24} & s_{44} & s_{45} & s_{46}  \\
    s_{15} & s_{25} & s_{45} & s_{55} & s_{56}  \\
    s_{16} & s_{26} & s_{46} & s_{56} & s_{66}  
 \end{bmatrix}
    \begin{bmatrix}
    \frac{\partial u_1}{\partial x_1}  \\
    \frac{\partial u_2}{\partial x_2}  \\
    \frac{\partial u_2}{\partial x_3}+\frac{\partial u_3}{\partial x_2}  \\
    \frac{\partial u_1}{\partial x_3}+\frac{\partial u_3}{\partial x_1}  \\
    \frac{\partial u_1}{\partial x_2}+\frac{\partial u_2}{\partial x_1}   
    \end{bmatrix}=    
    \begin{bmatrix}
    0 \\ 0 
    \end{bmatrix}\,.
\end{equation}

\begin{remark}
It should be noted that the presence of fibers may change the symmetry class of a material. For example, in the presence of straight fibers along the $x_3$-axis one cannot assume an isotropic solid. For a compressible isotropic solid
\begin{equation} 
\mathbf{c}=
\begin{bmatrix}
c_{11} & c_{12} & c_{12} & 0 & 0 & 0 \\
 c_{12} & c_{11} & c_{12} & 0 & 0 & 0 \\
 c_{12} & c_{12} & c_{11} & 0 & 0 & 0 \\
 0 & 0 & 0 & \frac{c_{11}-c_{12}}{2} & 0 & 0 \\
 0 & 0 & 0 & 0 & \frac{c_{11}-c_{12}}{2} & 0 \\
 0 & 0 & 0 & 0 & 0 & \frac{c_{11}-c_{12}}{2}
  \end{bmatrix}\,.
 \end{equation}
The constraint $\epsilon_{33}=0$ implies that $c_{12}(\sigma_{11}+\sigma_{22})+c_{11}\sigma_{33}=0$, and hence $c_{11}=c_{12}=0$, i.e., the compliance matrix identically vanishes. This implies that the fiber-reinforced solid is rigid, which is clearly not true as it can deform normal to the fiber direction, i.e., parallel to the $(x_1,x_2)$-plane. The fiber direction, here the $x_3$-axis, is a material preferred direction, i.e., the fiber-reinforced solid is not isotropic.
For the other seven symmetry classes we will assume that the fiber direction respects the corresponding symmetry in the sense that fibers are normal to one of the symmetry planes.
It should be noted that a triclinic solid reinforced with a family of fibers in any direction remains triclinic.
\end{remark}

\section{Universal displacements}  \label{Sec:Universal-Displacements}

In this section we consider all the possible seven symmetry classes: triclinic, monoclinic, tetragonal, trigonal, orthotropic, transversely isotropic, and cubic \citep{cowin1995anisotropic,Chadwick2001,Ting1996,ting2003generalized,cowin2007tissue}.
In order to determine the corresponding universal displacements, for each symmetry class the two equilibrium equations in \eqref{Universality-Inextensible} must hold for the arbitrary independent elastic constants. For each class we start with the following displacement field (recall that $\epsilon_{33}=0$)
\begin{equation} \label{Displacement-Field}
	\mathbf{u}(x_1,x_2,x_3)=(u_1(x_1,x_2,x_3),u_2(x_1,x_2,x_3),u_3(x_1,x_2))
	\,,
\end{equation}
and find the corresponding universality constraints, which are a set of linear PDEs that determine the universal displacements.

\subsection{Fiber-reinforced triclinic linear elastic solids} 

Triclinic solids are the least symmetric. Assuming reinforcement with fibers parallel to the $x_3$-axis in a Cartesian coordinate system $(x_1,x_2,x_3)$, the two equilibrium equations \eqref{Universality-Inextensible} must be satisfied for the $15$ arbitrary elastic constants. They give the following universality constraints for the displacement field \eqref{Displacement-Field}:
\begin{equation} \label{Universality-Triclinic1}
\begin{aligned}
	&  u_{1,11}(x_1,x_2,x_3)=0\,,\\
	&  u_{1,12}(x_1,x_2,x_3)=0\,,\\
	&  u_{1,13}(x_1,x_2,x_3)=0\,,\\
	&  u_{1,33}(x_1,x_2,x_3)=0\,,\\
	&  u_{2,12}(x_1,x_2,x_3)=0\,,\\
	&  u_{2,23}(x_1,x_2,x_3)=0\,,\\
	&  u_{2,22}(x_1,x_2,x_3)=0\,,\\
	&  u_{2,33}(x_1,x_2,x_3)=0\,,
\end{aligned}
\end{equation}
and
\begin{equation} \label{Universality-Triclinic2}
\begin{aligned}
 	&  u_{1,22}(x_1,x_2,x_3)+u_{2,12}(x_1,x_2,x_3)=0\,, \\
	&  2u_{1,12}(x_1,x_2,x_3)+u_{2,11}(x_1,x_2,x_3)=0\,,\\
	&  u_{1,11}(x_1,x_2,x_3)+u_{2,11}(x_1,x_2,x_3)=0\,,\\
	&  u_{1,22}(x_1,x_2,x_3)+2u_{2,12}(x_1,x_2,x_3)=0\,,\\
	&  u_{1,13}(x_1,x_2,x_3)+u_{3,11}(x_1,x_2)=0\,,\\
	&  u_{1,23}(x_1,x_2,x_3)+u_{3,12}(x_1,x_2)=0\,,\\
	&  2u_{1,13}(x_1,x_2,x_3)+u_{3,11}(x_1,x_2)=0\,,\\
	&  u_{2,13}(x_1,x_2,x_3)+u_{3,12}(x_1,x_2)=0\,,\\
	&  u_{2,23}(x_1,x_2,x_3)+u_{3,22}(x_1,x_2)=0\,,\\
	&  2u_{2,23}(x_1,x_2,x_3)+u_{3,22}(x_1,x_2)=0\,,\\
	&  2u_{1,23}(x_1,x_2,x_3)+u_{2,13}(x_1,x_2,x_3)+u_{3,12}(x_1,x_2)=0\,,\\
	&  u_{1,23}(x_1,x_2,x_3)+2u_{2,13}(x_1,x_2,x_3)+u_{3,12}(x_1,x_2)=0 \,.
\end{aligned}
\end{equation}
It is straightforward to show that the above universality constraints only admit homogeneous displacements. 
Thus, we have proved the following result.

\begin{prop}For a body made of a homogeneous triclinic material reinforced with inextensible fibers parallel to the $x_3$-axis, the universal displacements are of the following form
\begin{equation} \label{Triclinic-Universal-Displacements}
\begin{dcases}
	u_1(x_1,x_2,x_3) =a_0+a_1x_1+a_2x_2+a_3x_3 \,,\\
	u_2(x_1,x_2,x_3) =b_0+b_1x_1+b_2x_2+b_3x_3\,,\\
	u_3(x_1,x_2) =c_0+c_1x_1+c_2x_2	\,,
\end{dcases}
\end{equation}
where $a_i$, $b_i$, and $c_i$ are constant parameters, i.e., homogeneous displacement fields with vanishing $\epsilon_{33}$ strain.
\end{prop}

\begin{remark}
The third equilibrium equation in \eqref{Equilibrium-Equations} for a generic universal displacement field given in \eqref{Triclinic-Universal-Displacements} reads $T_{,3}(x_1,x_2,x_3)=0$, and hence $T=T(x_1,x_2)$, i.e., each fiber has a uniform tension. 
Suppose a fiber is normal to the boundary $\partial \mathcal{B}$ at $\mathbf{x}\in\partial \mathcal{B}$. At this point the traction boundary condition reads $\boldsymbol{\sigma}\mathbf{n}=(\sigma_{13},\sigma_{23},\sigma_{33})=(\sigma_{13},\sigma_{23},T)=(\bar{t}_1,\bar{t}_2,\bar{t}_3)$. This means that $T=\bar{t}_3$ is the axial force of the fiber everywhere. If this fiber intersects the boundary at another point $\mathbf{x}'\in\partial \mathcal{B}$ with unit normal along the fiber the traction boundary condition there must be consistent with that at $\mathbf{x}$, i.e., ${\bar{t}}^{\,'}_3=\pm\bar{t}_3$ (see also  \citep{Hayes1975}).
\end{remark}

\begin{remark}
The universal displacements of compressible triclinic solids are the set of all homogeneous displacements \citep{Yavari2020}.
In the case of fiber-reinforced solids, the inextensibility constraint forces the $u_3$ displacement to be a  function of only $(x_1,x_2)$, and hence, the set of universal displacements \eqref{Triclinic-Universal-Displacements} is a subset of the universal displacements of compressive triclinic solids.
\end{remark}

\subsection{Fiber-reinforced monoclinic linear elastic solids} 

A monoclinic solid at every point has a plane of reflection symmetry. Let us assume that everywhere the plane of symmetry is normal to the $x_3$-axis in a Cartesians coordinate system $(x_1,x_2,x_3)$.
We now assume that this body is reinforced by a family of inextensible fibers. 
In order to preserve this symmetry, let us assume that fibers are everywhere normal to the plane of symmetry, i.e., fibers are parallel to the $x_3$-axis. If the fibers are in any other direction, the reinforced solid becomes triclinic, a case that was discussed in the previous subsection.
A compressible monoclinic solid without reinforcement has $13$ independent compliance constants with the following compliance matrix:
\begin{equation}
    \mathbf{c}=\begin{bmatrix}
    c_{11} & c_{12} & c_{13} & c_{14} & 0 & 0 \\
    c_{12} & c_{22} & c_{23} & c_{24} & 0 & 0  \\
    c_{13} & c_{23} & c_{33} & c_{34} & 0 & 0  \\
    c_{14} & c_{24} & c_{26} & c_{44} & 0 & 0  \\
    0 & 0 & 0 & 0 & c_{55} & c_{56}  \\
    0 & 0 & 0 & 0 & c_{56} & c_{66}  
    \end{bmatrix}\,.
\end{equation}
The reduced compliance matrix has the following form
\begin{equation}
    \bar{\mathbf{c}}=\begin{bmatrix}
    c_{11} & c_{12} &  c_{14} & 0 & 0 \\
    c_{12} & c_{22} &  c_{24} & 0 & 0  \\
    c_{14} & c_{24} &  c_{44} & 0 & 0  \\
    0 & 0 & 0 & c_{55} & c_{56}  \\
    0 & 0 & 0 & c_{56} & c_{66}  
    \end{bmatrix}\,,
\end{equation}
which is a block diagonal matrix.
The reduced stiffness matrix $\bar{\mathbf{s}}=\bar{\mathbf{c}}^{-1}$ has the following block diagonal form
\begin{equation}
    \bar{\mathbf{s}}=\begin{bmatrix}
    s_{11} & s_{12} &  s_{14} & 0 & 0 \\
    s_{12} & s_{22} &  s_{24} & 0 & 0  \\
    s_{14} & s_{24} &  s_{44} & 0 & 0  \\
    0 & 0 & 0 & s_{55} & s_{56}  \\
    0 & 0 & 0 & s_{56} & s_{66}  
    \end{bmatrix}\,,
\end{equation}
where
\begin{equation}
    \begin{bmatrix}
    s_{11} & s_{12} &  s_{14}  \\
    s_{12} & s_{22} &  s_{24}   \\
    s_{14} & s_{24} &  s_{44} 
    \end{bmatrix}=\begin{bmatrix}
    c_{11} & c_{12} &  c_{14}  \\
    c_{12} & c_{22} &  c_{24}   \\
    c_{14} & c_{24} &  c_{44} 
    \end{bmatrix}^{-1}
    \,,\qquad
    \begin{bmatrix}
    s_{55} & s_{56}  \\
    s_{56} & s_{66}  
    \end{bmatrix}=\begin{bmatrix}
    c_{55} & c_{56}  \\
    c_{56} & c_{66}  
    \end{bmatrix}^{-1}\,. 
\end{equation}
It is seen that for the fiber-reinforced solid the number of independent elastic constants is reduced to $9$.  The two equilibrium equations \eqref{Universality-Inextensible} must be satisfied for the $9$ arbitrary elastic constants and give the following universality constraints for the displacement field \eqref{Displacement-Field}:
\begin{equation} \label{Universality-Monoclinic}
\begin{aligned}
	&  u_{1,12}(x_1,x_2,x_3)=0\,,\\
	&  u_{1,11}(x_1,x_2,x_3)=0\,,\\
	&  u_{1,33}(x_1,x_2,x_3)=0\,,\\
	&  u_{2,12}(x_1,x_2,x_3)=0\,,\\
	&  u_{2,22}(x_1,x_2,x_3)=0 \,, \\
	&  u_{2,33}(x_1,x_2,x_3)=0\,, \\
	&  u_{1,12}(x_1,x_2,x_3)+u_{2,11}(x_1,x_2,x_3)=0\,,\\
	&  u_{1,22}(x_1,x_2,x_3)+u_{2,12}(x_1,x_2,x_3)=0\,,\\
	&  u_{1,22}(x_1,x_2,x_3)+2 u_{2,12}(x_1,x_2,x_3)=0\,,\\
	&  2 u_{1,12}(x_1,x_2,x_3)+u_{2,11}(x_1,x_2,x_3)=0\,.
\end{aligned}
\end{equation}
It is straightforward to show that the above system of universality constraints admit the following family of universal displacements.

\begin{prop}
For a body made of a homogeneous monoclinic material with planes of symmetry parallel to the $(x_1,x_2)$-plane and reinforced with inextensible fibers parallel to the $x_3$-axis, the universal displacements are of the form
\begin{equation} \label{Monoclinic-Universal-Displacements}
\begin{dcases}
	u_1(x_1,x_2,x_3) =a_0+a_1x_1+a_2x_2+a_3x_3+a_{13}x_1x_3+a_{23}x_2x_3 \,,\\
	u_2(x_1,x_2,x_3) =b_0+b_1x_1+b_2x_2+b_3x_3+b_{13}x_1x_3+b_{23}x_1x_3\,,\\
	u_3(x_1,x_2,x_3) =u(x_1,x_2) 
	\,,
\end{dcases}
\end{equation}
where $u(x_1,x_2)$ is an arbitrary differentiable function, and $a_i$, $a_{ij}$, $b_i$, and $b_{ij}$ are arbitrary constants.
\end{prop}

\begin{remark}
The third equilibrium equation in \eqref{Equilibrium-Equations} for a generic universal displacement field given in \eqref{Monoclinic-Universal-Displacements} reads 
\begin{equation} 
	T_{,3}(x_1,x_2,x_3)
	+s_{55}   \left[u_{,11}(x_1,x_2)+a_{13}\right]
	+s_{45} \left[2u_{,12}(x_1,x_2)+a_{23}+b_{13}\right]
	+s_{44} \left[u_{,22}(x_1,x_2)+b_{23}\right]	=0
	\,.
\end{equation}
\end{remark}

\begin{remark}
The set of universal displacements \eqref{Monoclinic-Universal-Displacements} includes those of compressible monoclinic solids that were given in \citep{Yavari2020}.
\end{remark}

\subsection{Fiber-reinforced tetragonal linear elastic solids} 

In a tetragonal solid at every point there are five planes of symmetry. Four of the symmetry planes are coplanar while the fifth one is normal to the other four. Let us assume that in a Cartesian coordinate system $(x_1,x_2,x_3)$ the fifth plane of symmetry has normal along the $x_3$-axis.
The first two planes of symmetry are parallel to the $(x_1,x_3)$ and $(x_2,x_3)$-planes. The other two planes of symmetry are related to the ones parallel to the $x_1x_3$-plane by $\pi/4$ and $3\pi/4$ rotations about the $x_3$ axis.
A compressible tetragonal solid has $6$ independent compliance constants with the following compliance matrix:
\begin{equation}
    \mathbf{c}=\begin{bmatrix}
    c_{11} & c_{12} & c_{13} & 0 & 0 & 0 \\
    c_{12} & c_{11} & c_{13} & 0 & 0 & 0  \\
    c_{13} & c_{13} & c_{33} & 0 & 0 & 0  \\
    0 & 0 & 0 & c_{44} & 0 & 0  \\
    0 & 0 & 0 & 0 & c_{44} & 0  \\
    0 & 0 & 0 & 0 & 0 & c_{66}  
   \end{bmatrix}\,.
\end{equation}
We assume fiber reinforcement along $x_3$-axis. 
The reduced compliance matrix has the following form
\begin{equation}
     \bar{\mathbf{c}}=\begin{bmatrix}
    c_{11} & c_{12}  & 0 & 0 & 0 \\
    c_{12} & c_{11}  & 0 & 0 & 0  \\
    0 & 0 & c_{44} & 0 & 0  \\
    0 & 0 & 0 & c_{44} & 0  \\
    0 & 0 & 0 & 0 & c_{66}  
    \end{bmatrix}\,,
\end{equation}
which is a block diagonal matrix.
The reduced stiffness matrix $\bar{\mathbf{s}}=\bar{\mathbf{c}}^{-1}$ has the following block diagonal form
\begin{equation}
     \bar{\mathbf{s}}=\begin{bmatrix}
    s_{11} & s_{12}  & 0 & 0 & 0 \\
    s_{12} & s_{11}  & 0 & 0 & 0  \\
    0 & 0 & s_{44} & 0 & 0  \\
    0 & 0 & 0 & s_{44} & 0  \\
    0 & 0 & 0 & 0 & s_{66}  
    \end{bmatrix}\,,
\end{equation}
where
\begin{equation}
    \begin{bmatrix}
    s_{11} & s_{12}   \\
    s_{12} & s_{11}  
    \end{bmatrix}=\begin{bmatrix}
    c_{11} & c_{12}   \\
    c_{12} & c_{11}  
    \end{bmatrix}^{-1}
    \,,\qquad s_{44}=\frac{1}{c_{44}}\,,\qquad s_{66}=\frac{1}{c_{66}}    \,. 
\end{equation}
It is observed that for the fiber-reinforced solid the number of independent elastic constants is reduced to $4$.
The two equilibrium equations \eqref{Universality-Inextensible} must be satisfied for the $4$ arbitrary elastic constants and give the following universality constraints for the displacement field \eqref{Displacement-Field}:
\begin{equation} \label{Universality-Tetragonal}
\begin{aligned}
	& u_{1,12}(x_1,x_2,x_3) =0\,,\\
	& u_{1,11}(x_1,x_2,x_3) =0\,,\\
	& u_{1,33}(x_1,x_2,x_3) =0\,,\\
	& u_{2,12}(x_1,x_2,x_3) =0\,,\\
	& u_{2,22}(x_1,x_2,x_3) =0\,,\\
	& u_{2,33}(x_1,x_2,x_3) =0\,,\\
	& u_{1,12}(x_1,x_2,x_3)+u_{2,11}(x_1,x_2,x_3) =0\,,\\
	& u_{1,22}(x_1,x_2,x_3)+u_{2,12}(x_1,x_2,x_3) =0\,.
\end{aligned}
\end{equation}

It is straightforward to show that the above system of universality constraints admit the following family of universal displacements.

\begin{prop}
For a body made of a homogeneous tetragonal material with the tetragonal axis parallel to the $x_3$-axis and reinforced with inextensible fibers parallel to the $x_3$-axis, the universal displacements are of the following form
\begin{equation} \label{Tetragonal-Universal-Displacements}
\begin{dcases}
	u_1(x_1,x_2,x_3) =a_0+a_1x_1+a_2x_2+a_3x_3+a_{13}x_1x_3+a_{23}x_2x_3 \,,\\
	u_2(x_1,x_2,x_3) =b_0+b_1x_1+b_2x_2+b_3x_3+b_{13}x_1x_3+b_{23}x_1x_3\,,\\
	u_3(x_1,x_2,x_3) =u(x_1,x_2) 
	\,,
\end{dcases}
\end{equation}
where $u(x_1,x_2)$ is an arbitrary differentiable function, and $a_i$, $a_{ij}$, $b_i$, and $b_{ij}$ are arbitrary constants.
\end{prop}

\begin{remark}
The third equilibrium equation in \eqref{Equilibrium-Equations} for a generic universal displacement field given in \eqref{Tetragonal-Universal-Displacements} reads 
\begin{equation} 
	T_{,3}(x_1,x_2,x_3)+	s_{55}\left[\nabla^2u(x_1,x_2)+a_{13}+b_{23}\right]=0	\,.
\end{equation}
\end{remark}

\begin{remark}
The set of universal displacements \eqref{Tetragonal-Universal-Displacements} includes those of compressible tetragonal solids that were given in \citep{Yavari2020}.
\end{remark}

\begin{remark}
The universal displacements of $x_3$-fiber-reinforced monoclinic and tetragonal solids are identical. However, the PDEs determining their tension fields are different.
\end{remark}

\subsection{Fiber-reinforced trigonal linear elastic solids} 

In a trigonal solid at every point there are three planes of symmetry with normals that lie in the same plane and are related by $\pi/3$ rotations; two of the symmetry planes are related to the third one by rotations about a fixed axis by $\pi/3$ and $-\pi/3$. Let us assume that the trigonal axis is the $x_3$-axis. A compressible trigonal solid has $6$ independent compliance components, and its compliance  matrix has the following representation:
\begin{equation}
    \mathbf{c}=\begin{bmatrix}
    c_{11} & c_{12} & c_{13} & 0 & c_{15} & 0 \\
    c_{12} & c_{11} & c_{13} & 0 & -c_{15} & 0  \\
    c_{13} & c_{13} & c_{33} & 0 & 0 & 0  \\
    0 & 0 & 0 & c_{44} & 0 & -2c_{15}  \\
    c_{15} & -c_{15} & 0 & 0 & c_{44} & 0  \\
    0 & 0 & 0 & -2c_{15} & 0 & 2(c_{11}-c_{12})  
    \end{bmatrix}\,.
\end{equation}
Let us assume fiber reinforcement along $x_3$-axis. 
The reduced compliance matrix has the following form
\begin{equation}
    \mathbf{c}=\begin{bmatrix}
    c_{11} & c_{12}  & 0 & c_{15} & 0 \\
    c_{12} & c_{11}  & 0 & -c_{15} & 0  \\
     0 & 0 & c_{44} & 0 & -2c_{15}  \\
    c_{15} & -c_{15}  & 0 & c_{44} & 0  \\
    0 & 0 & -2c_{15} & 0 & 2(c_{11}-c_{12})  
    \end{bmatrix}\,.
\end{equation}
The reduced stiffness matrix $\bar{\mathbf{s}}=\bar{\mathbf{c}}^{-1}$ has the following representation
\begin{equation}
    \bar{\mathbf{s}}=\begin{bmatrix}
    s_{11} & s_{12} &  0 & s_{15} & 0 \\
    s_{12} & s_{11} &  0 & -s_{15} & 0  \\
    0 & 0 &  s_{44} & 0 & -s_{15}  \\
    s_{15} & -s_{15} &  0 & s_{44} & 0  \\
    0 & 0 & -s_{15} & 0 & \frac{1}{2}(s_{11}-s_{12})  
    \end{bmatrix}\,.
\end{equation}
It is seen that for the fiber-reinforced solid the number of independent elastic constants is reduced to $4$.
The two equilibrium equations \eqref{Universality-Inextensible} must be satisfied for the $4$ arbitrary elastic constants and give the following universality constraints for the displacement field \eqref{Displacement-Field}:
\begin{equation} \label{Universality-Trigonal}
\begin{aligned}
	& u_{1,33}(x_1,x_2,x_3) =0\,,\\
	& u_{2,33}(x_1,x_2,x_3) =0\,,\\
	&  u_{2,12}(x_1,x_2,x_3)-u_{1,22}(x_1,x_2,x_3)=0\,,\\
	&  u_{1,12}(x_1,x_2,x_3)-u_{2,11}(x_1,x_2,x_3)=0\,,\\
	&  u_{1,22}(x_1,x_2,x_3)+2u_{1,22}(x_1,x_2,x_3)+u_{2,12}(x_1,x_2,x_3)=0\,,\\
	& u_{1,12}(x_1,x_2,x_3)+2 u_{2,22}(x_1,x_2,x_3)+u_{2,11}(x_1,x_2,x_3)=0\,,\\
	&  u_{1,23}(x_1,x_2,x_3)+u_{2,13}(x_1,x_2,x_3)+u_{3,12}(x_1,x_2)=0\,,\\
	&  2 u_{1,13}(x_1,x_2,x_3)-2 u_{2,23}(x_1,x_2,x_3)-u_{3,22}(x_1,x_2)
	+u_{3,11}(x_1,x_2)=0\,.
\end{aligned}
\end{equation}
The first two PDEs imply that

\begin{equation} 
	u_1(x_1,x_2,x_3)=\xi_1(x_1,x_2)x_3+\xi_0(x_1,x_2)\,,\quad
	u_2(x_1,x_2,x_3)=\eta_1(x_1,x_2)x_3+\eta_0(x_1,x_2)\,.
\end{equation}
From \eqref{Universality-Trigonal}$_{3-4}$ one concludes that  
\begin{equation} 
	\eta_{0,12}=\xi_{0,22}\,,\quad \eta_{0,11}=\xi_{0,12}\,,\quad
	\eta_{1,12}=\xi_{1,22}\,,\quad \eta_{1,11}=\xi_{1,12}\,.
\end{equation}
Using these relations and \eqref{Universality-Trigonal}$_{5-6}$, it is concluded that $\nabla^2\xi_0=\nabla^2\xi_1=\nabla^2\eta_0=\nabla^2\eta_1=0$. Using these and the last two universality constraints in \eqref{Universality-Trigonal}, it can be shown that $\nabla^2u_3=c_0$, a constant.

\begin{prop}
For a body made of a homogeneous trigonal material with the trigonal axis parallel to the $x_3$-axis and reinforced with inextensible fibers parallel to the $x_3$-axis, the universal displacements are of the form
\begin{equation} \label{Trigonal-Universal-Displacements}
\begin{dcases}
	u_1(x_1,x_2,x_3) =\xi_0(x_1,x_2)+\xi_1(x_1,x_2)\,x_3 \,,\\
	u_2(x_1,x_2,x_3) =\eta_0(x_1,x_2)+\eta_1(x_1,x_2)\,x_3 \,,\\
	u_3(x_1,x_2,x_3) =\hat{u}(x_1,x_2) \,,
\end{dcases}
\end{equation}
where $u(x_1,x_2)$ is any constant-Laplacian function, $\xi_0$, $\xi_1$, $\eta_0$, and $\eta_1$ are harmonic functions such that $\eta_{0,12}=\xi_{0,22}$, $\eta_{0,11}=\xi_{0,12}$, $\eta_{1,12}=\xi_{1,22}$, and $ \eta_{1,11}=\xi_{1,12}$.
\end{prop}

\begin{remark}
The third equilibrium equation in \eqref{Equilibrium-Equations} for a generic universal displacement field given in \eqref{Trigonal-Universal-Displacements} reads 
\begin{equation} 
	T_{,3}(x_1,x_2,x_3)+	s_{44}\left[\xi_{1,1}(x_1,x_2)+\eta_{1,1}(x_1,x_2)+c_0\right]
	-4s_{15}\left[\xi_{0,22}(x_1,x_2)+x_3\xi_{1,22}(x_1,x_2)\right]=0	\,.
\end{equation}
\end{remark}

\begin{remark}
The universal displacements \eqref{Trigonal-Universal-Displacements} includes those of compressible trigonal solids that were given in \citep{Yavari2020}.
\end{remark}

\subsection{Fiber-reinforced orthotropic linear elastic solids} 

An orthotropic solid at every point has three mutually orthogonal symmetry planes. Let us assume that these are normal to the coordinate axes in a Cartesian coordinate system $(x_1,x_2,x_3)$.
A compressible orthotropic solid without reinforcement has $9$ independent compliance components, and its compliance matrix has the following representation:
\begin{equation}
    \mathbf{c}=\begin{bmatrix}
    c_{11} & c_{12} & c_{13} & 0 & 0 & 0 \\
    c_{12} & c_{22} & c_{23} & 0 & 0 & 0  \\
    c_{13} & c_{23} & c_{33} & 0 & 0 & 0  \\
    0 & 0 & 0 & c_{44} & 0 & 0  \\
    0 & 0 & 0 & 0 & c_{55} & 0  \\
    0 & 0 & 0 & 0 & 0 & c_{66}  
     \end{bmatrix}\,.
\end{equation}
In order to preserve the symmetry, we assume that fiber reinforcement is along one of the material preferred directions. Without loss of generality, let us assume that the fibers are parallel to the $x_3$-axis.
The reduced compliance matrix has the following form
\begin{equation}
    \bar{\mathbf{c}}=\begin{bmatrix}
    c_{11} & c_{12}  & 0 & 0 & 0 \\
    c_{12} & c_{22}  & 0 & 0 & 0  \\
    0 & 0 & c_{44} & 0 & 0  \\
    0 & 0 & 0 & c_{55} & 0  \\
    0 & 0 & 0 & 0 & c_{66}  
    \end{bmatrix}\,,
\end{equation}
which is a block diagonal matrix.
The reduced stiffness matrix $\bar{\mathbf{s}}=\bar{\mathbf{c}}^{-1}$ has the following block diagonal form
\begin{equation}
    \bar{\mathbf{s}}=\begin{bmatrix}
    s_{11} & s_{12}  & 0 & 0 & 0 \\
    s_{12} & s_{22}  & 0 & 0 & 0  \\
    0 & 0 & s_{44} & 0 & 0  \\
    0 & 0 & 0 & s_{55} & 0  \\
    0 & 0 & 0 & 0 & s_{66}  
    \end{bmatrix}\,,
\end{equation}
where
\begin{equation}
    \begin{bmatrix}
    s_{11} & s_{12}   \\
    s_{12} & s_{22}  
    \end{bmatrix}=\begin{bmatrix}
    c_{11} & c_{12}   \\
    c_{12} & c_{22}  
    \end{bmatrix}^{-1}
    \,,\qquad s_{44}=\frac{1}{c_{44}}\,,\qquad s_{55}=\frac{1}{c_{55}}\,,\qquad 
    s_{66}=\frac{1}{c_{66}}    \,. 
\end{equation}
It is seen that for the fiber-reinforced solid the number of independent elastic constants is reduced to $6$. The two equilibrium equations \eqref{Universality-Inextensible} must be satisfied for the $6$ arbitrary elastic constants and give the following universality constraints for the displacement field \eqref{Displacement-Field}:
\begin{equation} \label{Universality-Orthotropic}
\begin{aligned}
	&  u_{1,12}(x_1,x_2,x_3),=0\,,\\
	&  u_{1,11}(x_1,x_2,x_3)=0\,,\\
	&  u_{1,33}(x_1,x_2,x_3)=0\,,\\
	&  u_{2,12}(x_1,x_2,x_3)=0\,,\\
	&  u_{2,22}(x_1,x_2,x_3)=0\,,\\
	&  u_{2,33}(x_1,x_2,x_3)=0\,,\\
	&  u_{1,12}(x_1,x_2,x_3)+u_{2,11}(x_1,x_2,x_3)=0\,,\\
	&  u_{1,22}(x_1,x_2,x_3)+u_{2,12}(x_1,x_2,x_3)=0\,.
\end{aligned}
\end{equation}
It is straightforward to show that the above system of universality constraints admit the following family of universal displacements.

\begin{prop}
For a body made of a homogeneous orthotropic material with planes of symmetry normal to the coordinate axes in a Cartesian coordinate system $(x_1,x_2,x_3)$, and reinforced with inextensible fibers parallel to the $x_3$-axis, the universal displacements are of the following form
\begin{equation} \label{Orthotropic-Universal-Displacements}
\begin{dcases}
	u_1(x_1,x_2,x_3) =a_0+a_1x_1+a_2x_2+a_3x_3+a_{13}\,x_1x_3+b_{23}\,x_2x_3 \,,\\
	u_2(x_1,x_2,x_3) =b_0+b_1x_1+b_2x_2+a_3x_3+b_{13}\,x_1x_3+c_{23}\,x_2x_3\,,\\
	u_3(x_1,x_2,x_3) =u(x_1,x_2) \,,
\end{dcases}
\end{equation}
where $u(x_1,x_1)$ is an arbitrary differentiable function, and $a_i$, $a_{ij}$, $b_i$, and $b_{ij}$ are arbitrary constants.
\end{prop}

\begin{remark}
The third equilibrium equation in \eqref{Equilibrium-Equations} for a generic universal displacement field given in \eqref{Monoclinic-Universal-Displacements} reads 
\begin{equation} 
	T_{,3}(x_1,x_2,x_3)+s_{44} \left[u_{3,22}(x_2,x_3)+b_{23}\right]
	+s_{55} \left[u_{3,11}(x_2,x_3)+a_{13}\right]=0	\,.
\end{equation}
\end{remark}

\begin{remark}
The set of universal displacements \eqref{Orthotropic-Universal-Displacements} includes those of compressible orthotropic solids that was given in \citep{Yavari2020}.
\end{remark}

\begin{remark}
The universal displacements of $x_3$-fiber-reinforced monoclinic, tetragonal, and orthotropic  solids are identical. However, the PDEs determining their tension fields are different.
\end{remark}

\subsection{Fiber-reinforced transversely isotropic linear elastic solids} 

A transversely isotropic solid at every point has an axis of symmetry such that planes normal to it are isotropy planes. Let us assume that the axis of transverse isotropy is the $x_3$-axis in a Cartesian coordinate system $(x_1,x_2,x_3)$.  A compressible transversely isotropic solid has $5$ independent compliance components, and its compliance matrix has the following form:
\begin{equation}
    \mathbf{c}=\begin{bmatrix}
    c_{11} & c_{12} & c_{13} & 0 & 0 & 0 \\
    c_{12} & c_{11} & c_{13} & 0 & 0 & 0  \\
    c_{13} & c_{13} & c_{33} & 0 & 0 & 0  \\
    0 & 0 & 0 & c_{44} & 0 & 0  \\
    0 & 0 & 0 & 0 & c_{44} & 0  \\
    0 & 0 & 0 & 0 & 0 & 2(c_{11}-c_{12})  
    \end{bmatrix}\,.
\end{equation}
We assume that the transversely isotropic body is reinforced with a family of inextensible fibers parallel to the $x_3$-axis. 
The reduced compliance matrix has the following representation 
\begin{equation}
    \bar{\mathbf{c}}=\begin{bmatrix}
    c_{11} & c_{12}  & 0 & 0 & 0 \\
    c_{12} & c_{11}  & 0 & 0 & 0  \\
    0 & 0 & c_{44} & 0 & 0  \\
    0 & 0 & 0 & c_{44} & 0  \\
    0 & 0 & 0 & 0 & 2(c_{11}-c_{12})  
    \end{bmatrix}\,,
\end{equation}
which is a block diagonal matrix.
The reduced stiffness matrix $\bar{\mathbf{s}}=\bar{\mathbf{c}}^{-1}$ has the following block diagonal form
\begin{equation}
    \bar{\mathbf{s}}=\begin{bmatrix}
    s_{11} & s_{12}  & 0 & 0 & 0 \\
    s_{12} & s_{11}  & 0 & 0 & 0  \\
    0 & 0 & s_{44} & 0 & 0  \\
    0 & 0 & 0 & s_{44} & 0  \\
    0 & 0 & 0 & 0 & s_{66}  
    \end{bmatrix}\,,
\end{equation}
where
\begin{equation}
    \begin{bmatrix}
    s_{11} & s_{12}   \\
    s_{12} & s_{22}  
    \end{bmatrix}=\begin{bmatrix}
    c_{11} & c_{12}   \\
    c_{12} & c_{22}  
    \end{bmatrix}^{-1}
    \,,\qquad s_{44}=\frac{1}{c_{44}}\,,\qquad s_{66}=\frac{1}{2}(s_{11} - c_{12})   \,. 
\end{equation}
It is seen that for the fiber-reinforced solid the number of independent elastic constants is reduced to $3$. The two equilibrium equations \eqref{Universality-Inextensible} must be satisfied for the $3$ arbitrary elastic constants and give the following universality constraints for the displacement field \eqref{Displacement-Field}:
\begin{equation} \label{Universality-Transversely-Isotropic}
\begin{aligned}
	&  u_{1,33}(x_1,x_2,x_3)=0\,,\\
	&  u_{2,33}(x_1,x_2,x_3)=0\,,\\
	&  u_{1,12}(x_1,x_2,x_3)-u_{2,11}(x_1,x_2,x_3)=0\,,\\
	&  u_{2,12}(x_1,x_2,x_3)-u_{1,22}(x_1,x_2,x_3)=0\,,\\
	&  u_{1,22}(x_1,x_2,x_3)+2 u_{1,11}(x_1,x_2,x_3)+u_{2,12}(x_1,x_2,x_3)=0\,,\\
	&  u_{1,12}(x_1,x_2,x_3)+2 u_{2,22}(x_1,x_2,x_3)+u_{2,11}(x_1,x_2,x_3)=0\,.
\end{aligned}
\end{equation}
The first two universality constraints imply that $u_1(x_1,x_2,x_3)=\xi_1(x_1,x_2)\,x_3+\eta_1(x_1,x_2)$, and  $u_2(x_1,x_2,x_3)=\xi_2(x_1,x_2)\,x_3+\eta_2(x_1,x_2)$. The remaining PDEs give us
\begin{equation} \label{TI-Properties}
\begin{aligned}
	&  \xi_{2,11}=\xi_{1,12}\,,\quad \xi_{2,12}=\xi_{1,22}\,,\quad
	\eta_{2,11}=\eta_{1,12}\,,\quad \eta_{2,12}=\eta_{1,22}\,,\\
	& \nabla^2\xi_1=\nabla^2\xi_2=\nabla^2\eta_1=\nabla^2\eta_2=0\,.
\end{aligned}
\end{equation}
Thus, we have the following result.

\begin{prop}
For a body made of a homogeneous transversely isotropic material with the plane of isotropy parallel to the $(x_1,x_2)$-plane in a Cartesian coordinate system $(x_1,x_2,x_3)$, and reinforced with inextensible fibers parallel to the $x_3$-axis, the universal displacements are of the form
\begin{equation} \label{TI-Universal-Displacements}
\begin{dcases}
	u_1(x_1,x_2,x_3) =\xi_1(x_1,x_2)\,x_3+\eta_1(x_1,x_2) \,,\\
	u_2(x_1,x_2,x_3) =\xi_2(x_1,x_2)\,x_3+\eta_2(x_1,x_2)\,,\\
	u_3(x_1,x_2,x_3) =u(x_1,x_2) \,,
\end{dcases}
\end{equation}
where $u(x_1,x_2)$ is an arbitrary differentiable function, and $\xi_1$, $\xi_2$, $\eta_1$, and $\eta_2$ have the properties given in \eqref{TI-Properties}.
\end{prop}

\begin{remark}
The third equilibrium equation in \eqref{Equilibrium-Equations} for a generic universal displacement field given in \eqref{TI-Universal-Displacements} reads 
\begin{equation} 
	T_{,3}(x_1,x_2,x_3)+s_{44} \left[\xi_{1,1}(x_1,x_2)+\xi_{2,1}(x_1,x_2)
	+\nabla^2u(x_1,x_2)\right]=0	\,.
\end{equation}
\end{remark}

\begin{remark}
The set of universal displacement components $u_1$ and $u_2$ in \eqref{TI-Universal-Displacements} include those of the compressible transversely isotropic solids given in \citep{Yavari2020}. For compressible solids $u_3(x_1,x_2,x_3)=c_3x_3+g(x_1,x_2)$, where $c_3$ is a constant and $g$ is a harmonic function.
\end{remark}

\subsection{Fiber-reinforced cubic linear elastic solids} 

A cubic solid at every point has nine planes of symmetry with normals parallel to the edges and face diagonals of a cube. A compressible cubic solid has $3$ independent elastic constants. In a Cartesian coordinate system $(x_1,x_2,x_3)$ with coordinate lines parallel to the edges of the cube the compliance matrix is written as
\begin{equation}
    \mathbf{c}=\begin{bmatrix}
    c_{11} & c_{12} & c_{12} & 0 & 0 & 0 \\
    c_{12} & c_{11} & c_{12} & 0 & 0 & 0  \\
    c_{12} & c_{12} & c_{11} & 0 & 0 & 0  \\
    0 & 0 & 0 & c_{44} & 0 & 0  \\
    0 & 0 & 0& 0 & c_{44} & 0  \\
    0 & 0 & 0 & 0 & 0 & c_{44}  
    \end{bmatrix}\,.
\end{equation}
Let us assume that the body is reinforced with one family of fibers parallel to the $x_3$-axis. 
The reduced compliance matrix has the following form
\begin{equation}
    \bar{\mathbf{c}}=\begin{bmatrix}
    c_{11} & c_{12}  & 0 & 0 & 0 \\
    c_{12} & c_{11}  & 0 & 0 & 0  \\
    0 & 0 & c_{44} & 0 & 0  \\
    0 & 0& 0 & c_{44} & 0  \\
    0 & 0 & 0 & 0 & c_{44}  
    \end{bmatrix}\,,
\end{equation}
which is a block diagonal matrix.
The reduced stiffness matrix $\bar{\mathbf{s}}=\bar{\mathbf{c}}^{-1}$ has the following block diagonal form
\begin{equation}
    \bar{\mathbf{s}}=\begin{bmatrix}
    s_{11} & s_{12}  & 0 & 0 & 0 \\
    s_{12} & s_{11}  & 0 & 0 & 0  \\
    0 & 0 & s_{44} & 0 & 0  \\
    0 & 0 & 0 & s_{44} & 0  \\
    0 & 0 & 0 & 0 & s_{44}  
    \end{bmatrix}\,,
\end{equation}
where
\begin{equation}
    \begin{bmatrix}
    s_{11} & s_{12}   \\
    s_{12} & s_{11}  
    \end{bmatrix}=\begin{bmatrix}
    c_{11} & c_{12}   \\
    c_{12} & c_{11}  
    \end{bmatrix}^{-1}
    \,,\qquad s_{44}=\frac{1}{c_{44}}   \,. 
\end{equation}
It is seen that for the fiber-reinforced solid the number of independent elastic constants is still $3$.
The universality constrains read
\begin{equation} \label{Universality-Cubic}
\begin{aligned}
	& u_{1,12}(x_1,x_2,x_3) =0\,,\\
	& u_{1,11}(x_1,x_2,x_3) =0\,,\\
	& u_{2,12}(x_1,x_2,x_3) =0\,,\\
	& u_{2,22}(x_1,x_2,x_3) =0\,,\\
	& u_{1,12}(x_1,x_2,x_3)+u_{2,33}(x_1,x_2,x_3)+u_{2,11}(x_1,x_2,x_3) =0\,,\\
	& u_{1,33}(x_1,x_2,x_3)+u_{1,22}(x_1,x_2,x_3)+u_{2,12}(x_1,x_2,x_3) =0\,.
\end{aligned}
\end{equation}
It is straightforward to show that the above system of universality constraints admit the following family of universal displacements.

\begin{prop}
For a body made of a homogeneous cubic material reinforced with inextensible fibers parallel to the $x_3$-axis, the universal displacements are of the following form
\begin{equation}  \label{Cubic-Universal-Displacements}
\begin{dcases}
	u_2(x_1,x_2,x_3) =(a_0+a_1x_3)x_1+\chi(x_2,x_3) \,,\\
	u_3(x_1,x_2,x_3) =(b_0+b_1x_3)x_2+\psi(x_1,x_3) \,, \\
	u_3(x_1,x_2,x_3) =u(x_1,x_2) \,,
\end{dcases}
\end{equation}
where $u(x_1,x_2)$ is an arbitrary differentiable function, $a_i$ and $b_i$ are constant, and $\chi$ and $\psi$ are arbitrary harmonic functions.
\end{prop}

\begin{remark}
The third equilibrium equation in \eqref{Equilibrium-Equations} for a generic universal displacement field given in \eqref{Cubic-Universal-Displacements} reads 
\begin{equation} 
	T_{,3}(x_1,x_2,x_3)+s_{44} \left[a_{13}+b_{23}+\nabla^2u(x_1,x_2)\right]=0	\,.
\end{equation}
\end{remark}

\begin{remark}
The universal displacement components $u_1$ and $u_2$ in \eqref{Cubic-Universal-Displacements} are included in those of compressible cubic solids that were given in \citep{Yavari2020}. The intersection of the sets of universal displacement component $u_3$ in compressible and fiber-reinforced cubic solids is the set of all harmonic functions of the two variables $(x_1,x_2)$.
\end{remark}

\begin{table}[ht]
\centering
\begin{tabular}[t]{lcc}
\hline \rule{0pt}{1.0\normalbaselineskip}
Symmetry Class &Compressible Solids & Fiber-Reinforced Solids \\ [1mm]
\hline \rule{0pt}{1.0\normalbaselineskip}
Triclinic 				& $21$   &  $15$ \\
Monoclinic			& $13$  & $9$ \\
Tetragonal 			& $6$  & $4$  \\
Trigonal 				& $6$   & $4$  \\
Orthotropic 			& $9$   & $6$   \\
Transversely Isotropic 	& $5$  & $3$  \\
Cubic 				& $3$  & $3$  \\ 
\hline
\end{tabular}
\caption{The number of independent elastic constants for each symmetry class for both compressible and fiber-reinforced solids.}
\label{Table:Number}
\end{table}

\section{Conclusions}  \label{Sec:Conclusions}

The universal displacements of fiber-reinforced anisotropic linear elastic solids were characterized. The fibers are assumed to be inextensible and this introduces an internal constraint. In the presence of internal constraints the number of independent compliance components, and consequently the number of independent elastic constants, is reduced. 
We assumed a uniform distribution of straight fibers. Choosing a Cartesian coordinate system with $x_3$-axis parallel to the fibers, the tension field of the fibers appears only in the third equilibrium equation, which determines the tension field for a given displacement field.
We first noted that such a fiber-reinforced solid cannot be isotropic. This leaves seven symmetry classes. For each symmetry class we assumed that the fiber direction is normal to a plane of symmetry in order to preserve the symmetry of the matrix material.
Table \ref{Table:Number} summarizes the number of independent elastic constants for each symmetry class for both compressible and fiber-reinforced solids.
Universal displacements can be maintained in the absence of body forces by applying only boundary tractions. This means that universal displacements satisfy the first two equilibrium equations (normal to the fiber direction), or more precisely Navier's equations, for arbitrary elastic constants of the given symmetry class. This gives a set of linear PDEs---the universality constraints. In addition to the universality constraints the universal displacements must satisfy the inextensibility constraint $\epsilon_{33}=0$. 
Clearly, for any symmetry class, the set of universal displacements explicitly depends on the choice of the fiber distribution, which is an input of the problem. The results of this paper are valid only when fibers are parallel straight lines.
The following is the summary of our observations:
\begin{itemize}[topsep=0pt,noitemsep, leftmargin=10pt]
\item Universal displacements of monoclinic solids are homogeneous. Hence, up to the inextensibility constraint the universal displacements of compressible and fiber-reinforced triclinic solids are the same. 
\item The set of universal displacements of fiber-reinforced monoclinic, tetragonal, and orthotropic solids are identical. However, the PDEs governing their tension fields are different.
\item The set of universal displacements of fiber-reinforced monoclinic, tetragonal, trigonal, and orthotropic solids include the set of universal displacements of the corresponding compressible solids.
\item For fiber-reinforced transversely isotropic solids the sets of universal displacement components normal to the fiber direction include those of compressible transversely isotropic solids. The intersection of the sets of universal $u_3$ displacements is the space of harmonic functions of the two variables $(x_1,x_2)$.
\item For fiber-reinforced cubic solids the sets of universal displacement components normal to the fiber direction are included in those of compressible transversely isotropic solids. The intersection of the sets of universal $u_3$ displacements is the space of harmonic functions of the two variables $(x_1,x_2)$.
\end{itemize}
Table \ref{Table:UD} summarizes our results.
A goal of this paper was to investigate how an internal constraint affects the set of universal displacements. We have observed that other than the triclinic and cubic solids (a fiber-reinforced solid with straight fibers cannot be isotropic) in the other five classes the presence of inextensible fibers enlarges the set of universal displacements.

\begin{table}[hbt!]
\renewcommand{\arraystretch}{1.0}
\begin{center}
\resizebox{\textwidth}{!}{
\begin{tabular}{|c|c|c|c|}
\hline \rule{0pt}{1.0\normalbaselineskip}
Symmetry Class & Compressible Solids & Fiber-Reinforced Solids   \\[2mm]
\hline \rule{0pt}{2.50\normalbaselineskip}
Triclinic & $\begin{dcases}
	u_1(x_1,x_2,x_3) =a_0+a_1x_1+a_2x_2+a_3x_3 \\
	u_2(x_1,x_2,x_3) =b_0+b_1x_1+b_2x_2+b_3x_3 \\
	u_3(x_1,x_2) =c_0+c_1x_1+c_2x_2+c_3x_3	
\end{dcases}$ 
	  &  $\begin{dcases}
	u_1(x_1,x_2,x_3) =a_0+a_1x_1+a_2x_2+a_3x_3 \\
	u_2(x_1,x_2,x_3) =b_0+b_1x_1+b_2x_2+b_3x_3 \\
	u_3(x_1,x_2) =c_0+c_1x_1+c_2x_2	\,,
\end{dcases}$ \\[8mm]
	\hline \rule{0pt}{2.50\normalbaselineskip}
Monoclinic & $\begin{dcases}
	u_1(x_1,x_2,x_3) =a_0+a_1x_1+a_2x_2+a_3x_3+a_{23}\,x_2x_3 \\
	u_2(x_1,x_2,x_3) =b_0+b_1x_1+b_2x_2+b_3x_3-a_{23}\,x_1x_3 \\
	u_3(x_1,x_2) =c_0+c_1x_1+c_2x_2+c_3x_3	
\end{dcases}$ 
  & $\begin{dcases}
	u_1(x_1,x_2,x_3) =a_0+a_1x_1+a_2x_2+a_3x_3+a_{13}x_1x_3+a_{23}\,x_2x_3 \\
	u_2(x_1,x_2,x_3) =b_0+b_1x_1+b_2x_2+b_3x_3+b_{13}x_1x_3+b_{23}\,x_1x_3 \\
	u_3(x_1,x_2,x_3) =u(x_1,x_2) \end{dcases}$  \\[8mm] 
\hline \rule{0pt}{2.50\normalbaselineskip}
Tetragonal & $\begin{dcases}
	u_1(x_1,x_2,x_3) =a_0+a_1x_1+a_2x_2+a_3x_3+a_{23}\,x_2x_3+a_{13}\,x_1x_3 \\
	u_2(x_1,x_2,x_3) =b_0+b_1x_1+b_2x_2+b_3x_3-a_{13}\,x_2x_3+b_{13}\,x_1x_3 \\
	u_3(x_1,x_2) =c_0+c_1x_1+c_2x_2+c_3x_3+g(x_1,x_2)	
\end{dcases}$
  & $\begin{dcases}
	u_1(x_1,x_2,x_3) =a_0+a_1x_1+a_2x_2+a_3x_3+a_{13}\,x_1x_3+a_{23}\,x_2x_3 \\
	u_2(x_1,x_2,x_3) =b_0+b_1x_1+b_2x_2+b_3x_3+b_{13}\,x_1x_3+b_{23}\,x_1x_3 \\
	u_3(x_1,x_2,x_3) =u(x_1,x_2) \end{dcases}$  \\[8mm] 
\hline \rule{0pt}{3.0\normalbaselineskip}
Trigonal & $\begin{dcases}
	u_1(x_1,x_2,x_3) =a_0+a_1x_1+a_2x_2+a_3x_3
	+a_{123}x_1x_2x_3+a_{12}\,x_1x_2+a_{13}\,x_1x_3+a_{23}\,x_2x_3 \\
	u_2(x_1,x_2,x_3) =b_0+b_1x_1+b_2x_2+b_3x_3
	+\frac{1}{2}(a_{12}+a_{123}\,x_3)(x_1^2-x_2^2)+b_{13}\,x_1x_3-a_{13}\,x_2x_3 \\
	u_3(x_1,x_2) =c_0+c_1x_1+c_2x_2+c_3x_3
	-a_{123}\,x_1^2x_2-(a_{23}+b_{13})x_1x_2+\frac{1}{3}a_{123}\,x_2^3-a_{13}(x_1^2-x_2^2)
\end{dcases}$ 
  & $\begin{dcases}
	u_1(x_1,x_2,x_3) =\xi_0(x_1,x_2)+\xi_1(x_1,x_2)\,x_3 \\
	u_2(x_1,x_2,x_3) =\eta_0(x_1,x_2)+\eta_1(x_1,x_2)\,x_3 \\
	u_3(x_1,x_2,x_3) =\hat{u}(x_1,x_2) 
\end{dcases}$  \\[10mm] 
\hline \rule{0pt}{2.50\normalbaselineskip}
Orthotropic & $\begin{dcases}
	u_1(x_1,x_2,x_3) =a_0+a_1x_1+a_2x_2+a_3x_3+a_{23}\,x_2x_3 \\
	u_2(x_1,x_2,x_3) =b_0+b_1x_1+b_2x_2+b_3x_3+b_{13}\,x_1x_3 \\
	u_3(x_1,x_2) =c_0+c_1x_1+c_2x_2+c_3x_3+c_{12}\,x_1x_2	
\end{dcases}$ 
  & $\begin{dcases}
	u_1(x_1,x_2,x_3) =a_0+a_1x_1+a_2x_2+a_3x_3+a_{13}\,x_1x_3+a_{23}\,x_2x_3 \\
	u_2(x_1,x_2,x_3) =b_0+b_1x_1+b_2x_2+a_3x_3+b_{13}\,x_1x_3+b_{23}\,x_2x_3 \\
	u_3(x_1,x_2,x_3) =u(x_1,x_2) 
\end{dcases}$  \\[8mm] 
\hline  \rule{0pt}{2.50\normalbaselineskip}
Transversely Isotropic & $\begin{dcases} 
	u_1(x_1,x_2,x_3)=c_1x_1+c_2x_2+cx_2x_3+x_3h_1(x_1,x_2)+k_1(x_1,x_2) \\
	u_2(x_1,x_2,x_3)=-c_2x_1+c_1x_2-cx_1x_3+x_3h_2(x_1,x_2)+k_2(x_1,x_2) \\
	u_3(x_1,x_2,x_3)=c_3x_3+g(x_1,x_2) 
	\end{dcases}$
  & $\begin{dcases}
	u_1(x_1,x_2,x_3) =\xi_1(x_1,x_2)\,x_3+\eta_1(x_1,x_2) \\
	u_2(x_1,x_2,x_3) =\xi_2(x_1,x_2)\,x_3+\eta_2(x_1,x_2) \\
	u_3(x_1,x_2,x_3) =u(x_1,x_2) 
\end{dcases}$  \\[8mm] 
\hline \rule{0pt}{3.50\normalbaselineskip}
Cubic & $\begin{dcases} 
	u_1(x_1,x_2,x_3)=a_{123}\,x_1(x_3^2-x_2^2)
	+a_{13}\,x_1x_3+a_{12}\,x_1x_2+d_1x_1+g_1(x_2,x_3) \\
	u_2(x_1,x_2,x_3)=a_{123}\,x_2(x_1^2-x_3^2)
	+b_{12}\,x_1x_2-a_{13}\,x_2x_3+d_2x_2+g_2(x_1,x_3) \\
	u_3(x_1,x_2,x_3)=a_{123}\,x_3(x_2^2-x_1^2)
	-b_{12}\,x_1x_3-a_{12}\,x_2x_3+d_3x_3+g_3(x_1,x_2)
	\end{dcases}$
  & $\begin{dcases}
	u_2(x_1,x_2,x_3) =(a_0+a_1x_3)x_1+\chi(x_2,x_3) \\
	u_3(x_1,x_2,x_3) =(b_0+b_1x_3)x_2+\psi(x_1,x_3)  \\
	u_3(x_1,x_2,x_3) =u(x_1,x_2) 
\end{dcases}$  \\[10mm] 
\hline
\end{tabular}}
\end{center}
\caption[]{Universal displacements of compressible and fiber-reinforced (reinforcement along the $x_3$-axis) anisotropic linear elastic solids. The functions $k_1$, $k_2$, $h_1$, $h_2$, $\xi_0$, $\xi_1$, $\eta_0$, $\eta_1$, and $g$ are harmonic, $\hat{u}$ has a constant Laplacian, and $u$ is differentiable. $a_i$, $b_i$, $c_i$, $a_{ij}$, $b_{ij}$, and $a_{123}$ are constants.}
\label{Table:UD}
\end{table}

\section*{Acknowledgments}

This work was supported by NSF -- Grant No. CMMI 1939901.

\bibliographystyle{plainnat}
\bibliography{ref}

\end{document}